\documentstyle[twoside,fleqn,espcrc2]{article}
\title{
The operator-product expansion away from euclidean region}
\author{\underline{Jan Fischer}\thanks{
Supported in part by GAAV and GACR (Czech Republic) under 
grant numbers A1010711 and 202/96/1616 respectively.} and  Ivo Vrko\v{c} \\[3mm]
{%\it 
Institute of Physics, 18221 Prague 8 (J.F.), and Mathematical Institute, 
11567 Prague 1 (I.V.), Academy of Sciences of the Czech Republic}}
\begin{document}
\begin{abstract}
The role of the operator-product expansion in QCD calculations is discussed. 
Approximating the two-point correlation function by several terms and 
assuming an upper bound on the truncation error along the euclidean ray, 
we consider two model situations to examine how the bound develops 
with increasing deflection from the euclidean ray towards the cut. We 
obtain explicit bounds on the truncation error and show that they worsen 
with the increasing deflection.  
This result does not support the conventional believe that the remainder 
is constant for all angles in the complex energy plane. Further refinements 
of the formalism are discussed. 
\end{abstract}

\maketitle
\setcounter{footnote}{0}
%%%%%%%%%%%%%%%%%%%%%%%%%%%%%%%%
\section{Introduction}
A standard way to investigate the interconnection between perturbative and 
non-perturbative effects in quantum chromodynamics is the 
operator-product expansion (OPE) 
\cite{Wilson}.
In the momentum space, %the expansion (\ref{OPEx}) 
it has the form (we put $y=0$)
\begin{equation}
{\rm i} \int {\rm d}x {\rm e}^{{\rm i}qx}A(x)B(0) \approx
\sum_{k} C_{k}(q){\cal O}_{k} ,
\label{OPEq}
\end{equation} 
where $q$ is the total four-momentum,   %of the system,
$q^{2}=s=-Q^2$. 
The coefficients $C_{k}(q)$ are ordered according to the
increasing exponent $k$ in $s^{-k}$. 

Unlike perturbative expansions (in powers of the coupling constant
$\alpha_{s}$), very little is known about the mathematics  
of the operator product 
expansion. The knowledge is not sufficient to give (\ref{OPEq})
precise mathematical meaning, not even is known the exact composition of
terms on the right hand side of (\ref{OPEq}) (in particular, whether
terms exponential in the variable $Q^{-1}$ have to be added to 
describe strong-interaction processes). 
There are reasons to believe that the series
is divergent \cite{Shif94}, but it is not known whether it is asymptotic 
to the function expanded. The large-order behaviour of its terms is not
known either. Last not least, the symbol $\approx$ in (\ref{OPEq}) is
understood differently in different contexts.

In spite of serious mathematical uncertainties, the operator product
expansion is of paramount importance for solving a number of practical 
problems in quantum chromodynamics. To calculate many measurable 
quantities in QCD, it is necessary to investigate the properties of 
the OPE in the complex $Q^2$ plane 
away from  the euclidean region, i.e., away from the semiaxis $Q^2>0$. 
This leads us, with regard to (\ref{OPEq}), to the problem of finding 
conditions under which a power expansion of the type 
\begin{equation}
f(1/Q^2) \approx \sum_{k}a_{k}(q)/Q^{2k} 
\label{approx}
\end{equation}
can be extended to angles away from the euclidean semiaxis in the $Q^2$ 
plane. But such an extension is a delicate problem requiring precise 
mathematical conditions, which are not known here. 

In practical applications, it is usually assumed that the properties of
the operator product expansion away from the ray $Q^2>0$ are the same 
as those along it, except for the cut $Q^2<0$ (the Minkowski region). 
Instead of making this simplifying technical assumption, we prefer 
to introduce some model assumptions which would free the scheme from 
{\it a priori} neglecting the influence of the cut on the truncation 
error along a ray, keep the model possibly close to real situations 
and, simultaneously, give the problem precise mathematical meaning. 

%%%%%%%%%%%%%%%%%%%%%%%%%%%%%%%%%%%%%%%%
\section{The model} 
I will discuss the problem on the example of the following model situation. 
 Let $F(s)$ be 
holomorphic in the complex $s$ plane cut along the positive semiaxis, 
with the possible exception of a bounded domain around the origin. Let 
the numbers $a_{k}$, $k=0,1,2,...n-1$, and a positive number $A_{n}$ 
exist such that the inequality
\begin{equation}
|F(s) - \sum_{k=0}^{n-1} a_{k}/(-s)^{k}| < A_{n}/|s|^n  
\label{V2}
\end{equation}
holds for all real $s$ smaller than a negative number. The problem 
is under what conditions this inequality can be continued to the 
complex plane, and what will be its form away from the original 
region of validity, i.e. %along rays 
away from the negative real semiaxis (euclidean region). 

This problem is of interest for QCD because 
it could, when solved, tell us how the inequality (\ref{V2}) 
develops when the ray along which the operator product expansion is 
studied departs from the euclidean region  towards the Minkowski one. 
We shall discuss the problem under two simplifying assumptions. 

Note first that the inequality (\ref{V2}) amounts to assuming that 
the $n$-th order remainder $R_{n}(-1/s)$,  
\begin{equation} 
R_{n}(z) = f(z)-\sum_{k=0}^{n-1}a_{k} z^{k} , 
\label{rem}
\end{equation}
tends to zero for $z \rightarrow 0$ as the $n$-th 
power of $z$ for at least one value of $n$. (Here, we use the notation 
$z \equiv {\rm e}^{{\rm i}\varphi}=-1/s$ and $f(z) \equiv F(s)$.) 
This assumption may appear as too optimistic; indeed, it is difficult to 
make a realiable estimate of the truncation error even in the euclidean 
region, because it is not known whether there is a limit of QCD in which 
the operator-product expansion becomes exact. 

A second assumption we make is that the logarithmic energy dependence of 
the expansion coefficients $a_{k}$ can be neglected. 

We do not demand, on the other hand, that the series on the left-hand 
side of (\ref{V2}) be convergent or asymptotic to $F(s)$: our 
approach is more general and can be applied whenever the remainder 
obeys (\ref{V2}) at least for one value of $n$.  

Assuming the bound (\ref{V2}), we observe how it varies with the 
deflection of the ray away from euclidean region. It is natural to 
expect that the solution to the problem depends on additional 
assumptions imposed on the function $F(s) \equiv f(z)$. In 
particular, the form of the dependence of the resulting bound is 
determined by the assumed character of the discontinuity along the cut. 

We consider two possible sets of such additional assumptions.

%%%%%%%%%%%%%%%%%%%%%%%%%%%%%%
\section{Results}
(i) In the first of them, we assume that $f(z)$ admits the integral
representation of the form  
\begin{equation}
f(z)=\int_{0}^{\infty}\frac{\rho(t)}{1+zt}{\rm d}t 
\label{BOrs1} 
\end{equation}
with $z$ complex and $\rho(t)$ nonnegative for $t \geq 0$,  
and that the moment 
\begin{equation}
a_{n}=\int_{0}^{\infty} t^n \rho(t){\rm d}t
\label{BOrs2}
\end{equation}
exists for at least one positive integer $n$. Then it follows that 
the remainders $R_{k}(z)$ as defined by (\ref{rem}),  
with $k=1,2,...,n$, are bounded by the inequalities 
\begin{equation}
|R_{k}(z)| \leq a_{k}|z|^k  
\label{BOrs5}
\end{equation}
and 
\begin{equation}
|R_{k}(z)| \leq a_{k}\, |z|^k /|\sin \varphi|
\label{Vr3} 
\end{equation}
for Re$\,z > 0$ and Re\,$z < 0$ respectively. We see that the 
denominator on the right-hand side of the inequality (\ref{Vr3}) 
produces a looser estimate along rays lying 
in the halfplane Re\,$z < 0$, where the cut is present. 

Note that the integral representation (\ref{BOrs1}) is not equivalent 
to the dispersion integral representation that is typical  
for the QCD two-point functions, because these quantities usually obey 
a once-subtracted dispersion relation, in which case the moments 
(\ref{BOrs2}) do not exist. Contrary to this, the approach based on 
(\ref{BOrs1}) and ({\ref{BOrs2}) consists in assuming that at least several
lowest moments $a_{n}$ are convergent integrals. This feature may be of 
interest in some model situations, which we do not consider here. To get 
closer to physics, we consider an alternative scheme below.  

(ii) In this second scheme we assume, instead of the above
conditions, that $f(z)$ is bounded by a constant $M$ inside a 
circle of radius $d$ in the cut $z$ plane. The resulting 
bound on the remainder has the following form: 
\begin{equation}
|R_{n}(r{\rm e}^{{\rm i}\varphi})| \leq M_{n} (r/d_{n})^{n(1 - 
|\varphi|/\pi)} .  
\label{Vrk3}
\end{equation}
Here, $M_n$ and $d_n$ are constants which related to $M$ and $d$. This 
inequality is a special consequence of a general theorem, which will 
be published separately \cite{FV}. The inequality (\ref{Vrk3}) gives an 
upper bound on the remainder $R_{n}(z)$ along every ray tending to the 
origin (i.e., to infinite energy), the estimate becoming worse with 
increasing deflection from the positive real semiaxis (euclidean region), 
i.e., with the ray approaching closer the cut.  

Since the estimate (\ref{Vrk3}) might seem rather loose (note that $n$
in the exponent on the right hand side is multiplied by an angle-dependent
factor that vanishes on the cut), it is perhaps worth mentioning that 
there are functions that saturate it. 
Details will be discussed in \cite{FV}. 

%%%%%%%%%%%%%%%%%%%%%%%%%%%%%%%%%%%%%%%
\section{Discussion}
The bounds obtained are related to the problem of finding error 
estimates for the QCD calculations, in which contour integrals 
of the type
\begin{equation}
6 \pi {\rm i} \oint_{|s|=m^2} (1-s/m^2)^k 
(1+2s/m_{\tau}^2)^l P(s)\frac{{\rm d}s}{m^2} 
\label{Rtau}
\end{equation}
occur (with $k$ and $l$ being nonnegative integers). One example is 
the determination of $\alpha_{s}(m)$ from the $\tau$ lepton hadronic 
width \cite{LeDP}; in this case, $m$ is the $\tau$-lepton mass, $k$=2, 
$l$=1, and $P(s)$ is a combination of the electromagnetic two-point 
correlation functions $\Pi(s)$ for the vector and the axial vector 
colour singlet massless quark currents, the $\Pi(s)$ being defined as 
\begin{eqnarray}
\nonumber
\Pi ^{\mu \nu} =(g^{\mu \nu}q^2 - q^{\mu}q^{\nu}) \Pi (-q^2) \\
\nonumber
= {\rm i} \int {\rm d}^4 x {\rm e}^{-{\rm i}qx} 
\langle 0 | \, {\rm T} (j^{\mu}(x) j^{\nu}(0)) \, | 0 \rangle  
\end{eqnarray}
for two currents $j^{\mu}$ and $j^{\nu}$. 
Note that the integral (\ref{Rtau}) may receive  
essential contributions from the region near $s=m^{2}$, 
where the OPE has little chance appropriately to represent the 
function expanded. 
A fortunate circumstance is that the double zero of the kinematic 
factor $(1-s/m^{2})^2$ in the integrand 
suppresses the contribution from this dangerous 
segment. But a quantitative analysis of this argument is, to the best 
of our knowledge, still lacking. 

It has been the motivation of the present research to find an explicit 
quantitative estimate of the truncation error. 
In considering the problem of the evaluation of the contour 
integrals of the type (\ref{Rtau}), 
we propose a model scheme of assumptions with the aim to evaluate the
influence of the cut $Q^2<0$ by estimating the truncation error along 
different rays in the complex $Q^2$ plane. 
In this scheme, the property (\ref{V2}) occupies the central position. 
Its combination with analyticity allows an extension of (\ref{V2}) 
into the complex $s$ plane.    

We have cosidered two possible additional assumption sets, (i) 
(\ref{BOrs1}) and (\ref{BOrs2}), and (ii) the assumption that $f(z)$ is
bounded inside a circle of the cut $z$ plane. While the case (i) leads 
to the bounds (\ref{BOrs5}) and ({\ref{Vr3}), the case (ii) yields the 
bound (\ref{Vrk3}), in which the exponent itself depends on the phase 
$\varphi$, thereby representing a significant worsening of the 
{\it asymptotic} bound along rays that are close to the cut.  

In case (i), an improvement of the integrated error in the 
integrals of the type (\ref{Rtau}) may be reached by reversing 
the order of the $s$-integration and the $z$-integration. No 
such chance seems to exist in the case (ii). 

Our results can be considered as a step out of the conventional 
scheme, in which %is used in applications \cite{LeDP} and is based on 
the pragmatic apriori assumption is made that the truncation error 
is the same along all 
rays except the cut. Our bounds are based on a plausible, but still crude 
and fictitious picture  
of the operator product expansion, possibly good for certain model 
situations. 
In application to physics, however, 
as was pointed out above, the scheme %used by us 
is not fully appropriate either, due to the simplifying assumptions %made. 
discussed in Section 2 of this talk. A further refinement is desirable. 
Next step should include a regard to the logarithmic $q$--dependence of 
the coefficients  $a_{k}(q)$. Work along these lines is in progress.   

\noindent{\bf Acknowledgements}

Helpful discussions on the subject with M. Beneke, I. Caprini, E. de Rafael, 
S. Narison, M. Neubert and J. Stern are gratefully acknowledged. I am 
indebted to A. de R\'{u}jula for hospitality at the CERN TH Division, 
and to Stephan Narison for inviting me to this inspiring conference.  

%%%%%%%%%%%%%%%%%%%%%%%%%%%%%%%%%%%%%%%%%%%%%


\begin{thebibliography}{99}

\bibitem{Wilson} K.G. Wilson, Phys.Rev. {\bf 179} (1969) 1499  

\bibitem{Shif94} M.A. Shifman, {\it in} Continuous Advances in QCD 1994,
A. Smilga (ed.), p. 249 (World Scientific, Singapore, 1994)
[hep-ph/9405246] 

\bibitem{FV} J. Fischer and I. Vrko\v{c}, to be published

\bibitem{Shif} M. Shifman, Int.J.Mod.Phys. {\bf A 11} (1996) 3195 

\bibitem{LeDP} E. Braaten, S. Narison and A. Pich, Nucl.Phys.
{\bf B 373} (1992) 581  

A. Pich: QCD tests from tau decays. Invited talk at the 20th
Johns Hopkins Workshop (Heidelberg, 27-29 June 1996), hep-ph/9701305 

F. LeDiberder and A. Pich, Phys.Lett. {\bf B 289} (1992) 165

\end{thebibliography}
\end{document}